\documentclass[preprint,prd,noshowpacs,nofootinbib]{revtex4}

\usepackage{color} 
\usepackage{amssymb}
\usepackage{amsmath}
\usepackage{graphicx} 
\usepackage{float}
\usepackage{braket}     
\usepackage{pdfpages}
\usepackage{appendix}
\usepackage{epstopdf}
\usepackage[utf8]{inputenc}
\usepackage{soul}

\begin{document}

\title{AC Stark effect or time-dependent Aharonov-Bohm effect for particle on a ring}

\author{Patrick Hinrichs}
\email{phinrichs92@mail.fresnostate.edu}
\affiliation{Department of Physics, California State University Fresno,  Fresno, CA 93740-8031, USA}

\author{Nader Inan}
\email{ninan@ucmerced.edu}
\affiliation{Clovis Community College, 10309 N. Willow, Fresno, CA 93730 USA}
\affiliation{University of California, Merced, School of Natural Sciences, 
Merced, CA 95344, USA}
\affiliation{Department of Physics, California State University Fresno, Fresno, CA 93740-8031, USA }

\author{Douglas Singleton}
\email{dougs@mail.fresnostate.edu}
\affiliation{Department of Physics, California State University Fresno, Fresno, CA 93740-8031, USA}

\date{\today}

\begin{abstract}
We study the effect of a time-varying solenoidal vector potential for a quantum particle confined to a ring. The setup appears to be a time-varying version of the Aharonov-Bohm effect, but since the particle moves in the presence of fields, it is not strictly an Aharonov-Bohm effect. The results are similar to the ac Stark effect, but with a time-varying electric field coming from the vector potential, rather than the scalar potential.  We compare and contrast the present effect with the standard ac Stark effect. The signature of this setup is the generation of quasi-energy sidebands which are observable via spectroscopy.
\end{abstract}

\maketitle

\section{Introduction}

A quantum ring is a simple quantum system in which a particle is confined to move on a ring of a fixed radius, $R$. For a particle of mass $m$ and charge $e$ the Hamiltonian of this system is 
\begin{equation}
    \label{hamiltonian0}
    H_0 = \frac{1}{2m} p_\varphi ^2 = - \frac{\hbar^2}{2mR^2} \frac{\partial ^2}{\partial \varphi^2}  ~,
\end{equation}
where the momentum operator in the $\varphi$-direction is $p_\varphi = \frac{-i \hbar}{R} \frac{\partial }{\partial \varphi}$. This system has no potential energy, but it possesses kinetic energy, with motion constrained to be on a one-dimensional ring. The normalized eigenfunctions and eigenenergies for the time-independent Schr{\"o}dinger equation, $H_0 \psi ^{(0)}_n = E^{(0)}_n \psi^{(0)}$, are easily found and given by 
\begin{equation}
    \label{wav-func}
    \psi^{(0)}_n = \frac{1}{\sqrt{2 \pi R}} e^{i n \varphi} ~~~~~{\rm and}~~~~~ E^{(0)}_n =  \frac{\hbar ^2 n^2}{2mR^2} ~,
\end{equation}
where $n$ is an integer. This system bears some similarity to a particle in a hard wall box, which also has sinusoidal wave functions and energy eigenvalues proportional to $n^2$. For quantum rings, this behavior comes from periodic boundary conditions instead of hard wall boundary conditions.  
 
A slight generalization of the quantum ring occurs if a magnetic flux $\Phi_0$ is threaded through the middle of the ring. This system is connected to the Aharonov-Bohm (AB) effect \cite{ab} as discussed in \cite{tong,griffiths}. If one has a constant magnetic field $B_0$ confined to a cylinder of radius $a <R$, the vector potential at the ring radius $R$ is $A_\varphi = \frac{\Phi_0}{2 \pi R}$, where $\Phi_0 = \pi a^2 B_0$ is the magnetic flux of the cylinder. Using this $A_\varphi$ and minimal coupling, $p_\varphi \to p_\varphi + e A_\varphi$, the Hamiltonian from \eqref{hamiltonian0} becomes  
\begin{equation}
    \label{hamiltonian}
    H = \frac{1}{2m} (p_\varphi + e A_\varphi)^2 = \frac{1}{2mR^2} \left( -i \hbar \frac{\partial }{\partial \varphi} + \frac{e \Phi_0}{2 \pi }\right )^2 ~.
\end{equation}
The time-independent Schr{\"o}dinger equation for this Hamiltonian ({\it i.e.} $H \psi_n = E_n \psi_n$) is solved by the same eigenfunctions used for the original Hamiltonian, $H_0$ ({\it i.e.} $\psi_n = \psi^{(0)}_n = \frac{1}{\sqrt{2 \pi R}} e^{i n \varphi} $).  
However, the energy eigenvalues are shifted to take the form  
\begin{equation}
    \label{energy}
    E_n =  \frac{1}{2mR^2} \left(\hbar n + \frac{e \Phi_0}{2 \pi }\right )^2 = \frac{\hbar ^2}{2mR^2} \left(n + \frac{\Phi_0}{\Phi_{QM}}\right )^2~,
\end{equation}
where $\Phi_{QM} = \frac{2\pi \hbar}{e}$ is the quantum of magnetic flux. There will be a detectable shift in the energy levels if the magnetic flux is not an integer multiple of $\Phi _{QM}$ (\textit{i.e.} $\Phi_0 \ne m \Phi _{QM}$, with $m$ being an integer).  On the other hand, if $\Phi_0 = m \Phi _{QM}$, the shift of the energy levels will not be detectable since they will simply shift from one level to another. For example, if $\Phi_0 = \Phi _{QM}$ ({\it i.e.} $m=1$) then the energy levels will shift from $E_n \propto n^2$ without magnetic flux, to $E_n \propto (n+1)^2$ with magnetic flux.  This will shift the original $n=...-1, 0, 1, ...$ states into the $n=...0, 1, 2...$ states, which will be undetectable, since the energy differences between levels remain the same. For these quantum rings threaded by static magnetic flux, the signal of the Aharonov-Bohm effect is a shift in energy levels (assuming  $\Phi_0 \ne m \Phi _{QM}$) rather than a shift in interference pattern as in the standard magnetic Aharonov-Bohm setup. 

In this work, we study the system given by the Hamiltonian in \eqref{hamiltonian} but for a magnetic flux that is sinusoidally time-varying. 
Although this system seems closely related to the Aharonov-Bohm effect, in the following section, we argue that it 
is more closely related to the ac Stark effect or the Autler-Townes effect \cite{autler}. 

The ac Stark effect \footnote{See the excellent review \cite{DK}, which we will rely on heavily in comparing our system of a quantum ring  threaded by a time-varying magnetic flux, with the standard ac Stark effect} has terms that are linear and quadratic in the electric field strength. As shown in equation (14) of \cite{DK}, the time-dependent Schr{\"o}dinger equation has two interaction terms: $-d F \cos (\omega t)$ and $-\frac{1}{2} \alpha F^2 \cos ^2 (\omega t)$, where $F$ is the amplitude of the electric field, $d$ is the constant dipole moment of the system, and $\alpha$ is the polarizability induced by the field. We will find that the quantum ring threaded by a sinusoidal magnetic flux will have a mathematically identical wavefunction and energy spectrum as the ac Stark effect, with the difference that for the present system the linear and quadratic terms are related to one another. 

\section{Time-dependent vector potential}

In this section, we provide details about the vector potential and fields for the solenoid with a sinusoidally time-varying magnetic flux. For a solenoid of radius $a$, with $n$ turns per unit length and a sinusoidally varying current of $I(t) = I_0 \cos (\omega t)$, the vector potential for $\rho >a$ ({\it i.e.} outside the solenoid) is given by \cite{zangwill} \cite{templin} \footnote{We have converted the SI units used in \cite{zangwill} \cite{templin} to cgs units, and we use cgs units throughout the paper.}
\begin{equation}
    \label{vec-pot}
    {\bf A} ({\bf x}, t) = \frac{2 \pi n I_0}{c} \pi a J_1 (k a)\left[ J_1(k \rho) \sin (\omega t) - Y_1 (k \rho) \cos (\omega t) \right] \boldsymbol{{\hat \varphi}}~.
\end{equation}
In equation \eqref{vec-pot}, $k = \omega /c$, $J_1$ is a Bessel function of order 1 and $Y_1$ is a Bessel function of the second kind of order $1$. From \eqref{vec-pot} the time-varying ${\bf A} ({\bf x}, t)$ generates time-varying electric and magnetic fields for $\rho >a$ of the form
\begin{equation}
    \label{e-fields}
    {\bf E} ({\bf x}, t) = - \frac{\partial {\bf A}}{\partial t} = - \frac{2 \pi n I_0 \omega}{c} \pi a J_1 (k a)\left[ J_1(k \rho) \cos (\omega t) + Y_1 (k \rho) \sin (\omega t) \right] {\boldsymbol{{\hat \varphi}}} ~,
\end{equation}
and 
\begin{equation}
    \label{b-fields}
    {\bf B} ({\bf x}, t) = \nabla \times {\bf A} = \frac{2 \pi n I_0 \omega}{c^2} \pi a J_1 (k a)\left[ J_0 (k \rho) \cos (\omega t) - Y_0 (k \rho) \sin (\omega t) \right] {\hat {\bf z}} ~.
\end{equation}

Since the electric and magnetic fields are non-zero {\it outside} the solenoid, this is not an Aharonov-Bohm effect which requires zero fields but non-zero potentials. Nevertheless, this is the same setup as the time-independent Aharonov-Bohm effect reviewed in the introduction, and as mentioned we will show that this setup is mathematically similar to the ac Stark effect or Autler-Townes effect \cite{autler}, with the difference that the linear and quadratic terms in our case are directly related to one another.  

The vector potential, ${\bf A}$, from \eqref{vec-pot}  is evaluated at the location of the radius of the quantum ring, $R = \rho$. For the Bessel functions in \eqref{vec-pot} we assume that the product of the wavenumber $k$ and the radii, $a$ and $R$, are small {\it  i.e.} $ka  < kR \ll 1$. Since $\omega = k c$, these limits are essentially low-frequency limits. This point will be revisited in Section III, where we consider realistic experimental parameters. Under these conditions we have the asymptotic expansions $J_1 (ka) \approx \frac{1}{\Gamma(2)} \left( \frac{ka}{2}\right) =\frac{ka}{2}$,  $J_1 (kR) \approx \frac{1}{\Gamma(2)} \left( \frac{kR}{2}\right) =\frac{kR}{2}$, and $Y_1 (kR) \approx - \frac{\Gamma(1)}{\pi} \frac{2}{kR} = - \frac{2}{\pi kR}$. Inserting these into \eqref{vec-pot} gives 
\begin{equation}
    \label{vec-pot2}
    {\bf A} ({\bf x} , t)  
    \approx \frac{2 \pi ^2 a^2n I_0}{c R} \left[ \frac{k^2 R^2}{4} \sin (\omega t) + \frac{1}{\pi} \cos (\omega t) \right] \boldsymbol{{\hat \varphi}} \approx  \frac{\Phi _0}{2 \pi R} \cos (\omega t) \boldsymbol{{\hat \varphi}}  ~.
\end{equation}
In \eqref{vec-pot2} we have written the flux carried by the solenoid as $\Phi_0 = B_0 \pi a^2$, with $B_0 = \frac{4 \pi n I_0}{c}$ being the magnetic field magnitude inside the solenoid. In the last step in \eqref{vec-pot2}, we have dropped terms of order ${\cal O} (k^2 R^2)$, which is in line with the condition $kR \ll 1$. In this approximation, the vector potential in \eqref{vec-pot2} is that of a static solenoid ({\it i.e.} $\frac{\Phi_0 }{2 \pi R}$) multiplied by the sinusoidal time dependence, $\cos (\omega t)$. Note that with the vector potential approximated in \eqref{vec-pot2}, the magnetic field outside the solenoid is zero ({\it i.e.} ${\bf B} = \nabla \times {\bf A} =0$), but the electric field is non-zero ({\it i.e.} ${\bf E} = - \partial_t {\bf A} = \frac{\Phi_0 \omega}{2 \pi R} \sin (\omega t)$). These match the results from \cite{templin} -- see equations (20) and (21). 

Using the vector potential from \eqref{vec-pot2} the time-independent Hamiltonian of \eqref{hamiltonian} becomes the time-dependent Hamiltonian of the form 
\begin{equation}
    \label{hamiltonian2}
    H = \frac{1}{2m} (p_\varphi + e A_\varphi)^2 = \frac{1}{2mR^2} \left( -i \hbar \frac{\partial }{\partial \varphi} + \frac{e \Phi_0 \cos (\omega t) }{2 \pi }\right )^2 ~.
\end{equation}
We now solve the time-dependent  Schr{\"o}dinger equation: $H \psi = i \hbar  \frac{ \partial \psi}{\partial t}$. The wavefunction will have a more complicated time part than for the case of the time-independent vector potential, which had the standard form $e^{-i E t / \hbar}$. The spatial part of the wavefunction is assumed to have the same form as in the time-independent case, namely $\propto e^{i n \varphi}$.
We now write the full wavefunction as $\psi ({\bf x}, t) = K e^{i n \varphi} e^{-i f(t)/\hbar}$, where $f(t)$ is some energy-like function to be determined, and $K$ is a normalization constant which turns out to be the same as in the static case, namely $K= \frac{1}{\sqrt{2 \pi R}}$. Applying \eqref{hamiltonian2} to this wavefunction and solving the time-dependent Schr{\"o}dinger equation for $f'(t) = \frac{df}{dt}$ yields 
\begin{equation}
    \label{energy-4}
f'(t) = \frac{\hbar ^2 }{2mR^2} \left( n^2 + \frac{2 n \Phi _0 \cos (\omega t)}{\Phi _{QM}} + \frac{\Phi^2 _0 \cos ^2 (\omega t)}{\Phi_{QM} ^2}\right)~.
\end{equation}
Equation \eqref{energy-4} can be integrated with respect to $t$ giving 
\begin{equation}
    \label{energy-5}
f(t) = \frac{\hbar ^2 }{2mR^2} \left( n^2 t + \frac{2 n \Phi _0 \sin (\omega t)}{\omega \Phi _{QM}} + \frac{\Phi^2 _0}{2 \Phi_{QM} ^2}\left[t + \frac{\sin(2 \omega t)}{2 \omega} \right] \right)~.
\end{equation}
In the limit when $\omega \to 0$, we recover the results of the previous section. 
\begin{equation}
    \label{energy-6}
\lim_{\omega \to 0} ~ \frac{\hbar ^2 }{2mR^2} \left( n^2 t + \frac{2 n \Phi _0 \sin (\omega t)}{\omega \Phi _{QM}} + \frac{\Phi^2 _0}{2 \Phi_{QM} ^2}\left[t + \frac{\sin(2 \omega t)}{2 \omega} \right] \right) \to  \frac{\hbar ^2 }{2mR^2} \left(n + \frac{\Phi_0}{\Phi_{QM}}\right) ^2 t ~.
\end{equation}
Looking at \eqref{energy} we can see that \eqref{energy-6} is just $E_n t$ as expected. 

In the case when $\omega \ne 0$, we can use the result in \eqref{energy-5} to write the wavefunction as
\begin{eqnarray}
    \label{wav-func-2}
    \psi ({\bf x}, t) &=& \frac{1}{\sqrt{2 \pi R}} e^{i n \varphi} e^{-i f(t)/\hbar} \nonumber \\
    &=& \frac{1}{\sqrt{2 \pi R}} e^{i n \varphi} \exp \left[\left( \frac{-i}{\hbar} \right)\frac{  \hbar ^2 }{2mR^2} \left( n^2 + \frac{\Phi^2 _0}{2 \Phi_{QM} ^2} \right)  t \right] \nonumber \\
    &\times&
    \exp \left[ \left( \frac{-i}{\hbar} \right) \frac{\hbar ^2 }{2mR^2} \left( \frac{2 n \Phi _0 \sin (\omega t)}{\omega \Phi _{QM}} + \frac{\Phi^2 _0\sin(2 \omega t)}{4 \omega \Phi_{QM} ^2 } \right)  \right]~.
\end{eqnarray}
The term in \eqref{wav-func-2} that is linear in $t$ is like the $e^{-i E_n t/\hbar}$ term in \eqref{energy}. However, in going from \eqref{energy} to \eqref{wav-func-2} $E_n$ is replaced by $E'_n$ according to 
\begin{equation}
\label{En'}
E_n= \frac{\hbar ^2 }{2mR^2} \left(n + \frac{\Phi_0}{\Phi_{QM}}\right )^2 ~~~\to~~~ E'_n = \frac{\hbar ^2 }{2mR^2} \left( n^2 + \frac{\Phi^2 _0}{2 \Phi_{QM} ^2} \right) ~.
\end{equation}
The energy shift given by the first expression, $E_n$, in \eqref{En'} is spectroscopically detectable as long as $\frac{\Phi _0}{\Phi_{QM}}$ is not an integer \cite{tong}. However, the energy shift for $E_n '$ from \eqref{En'} is not spectroscopically detectable regardless of the value of $\frac{\Phi _0^2}{2 \Phi_{QM}^2}$, since this provides the same shift to all base energies $\frac{\hbar ^2 n^2}{2mR^2}$. Below, we will show that the energies $E_n '$ in \eqref{En'} develop energy sidebands that {\it are} spectroscopically detectable. These sidebands are similar to those that appear in the ac Stark effect \cite{DK}.

The two sinusoidal terms in the last line of \eqref{wav-func-2} are worked out in Appendix A using the Jacobi-Anger expansion. From \eqref{ja-5}, we can write the wave function from \eqref{wav-func-2} as
\begin{equation}
    \label{wav-func-2a}
    \psi ({\bf x}, t)  = \frac{e^{i n \varphi}}{\sqrt{2 \pi R}} 
    \sum _{r, s =-\infty} ^{\infty} (-1)^{r} J_{r+2s} \left( \frac{n\hbar \Phi_0}{m R^2 \omega \Phi _{QM}} \right) J_s \left( \frac{\hbar \Phi ^2_0}{8 m R^2 \omega \Phi ^2 _{QM}} \right) e^{-i (r \hbar \omega + E_n ')t/ \hbar} ~.
\end{equation}
We want to compare the result in \eqref{wav-func-2a} against the known result for the ac Stark effect found in the review \cite{DK}. To do this, we define $C_{r}$ as
\begin{eqnarray}
    \label{Ck}
    C_{r} &=&  \sum _{s =-\infty} ^{\infty} (-1)^{r} J_{r+2s} \left( \frac{n\hbar \Phi_0}{m R^2 \omega \Phi _{QM}} \right) J_s \left( \frac{\hbar \Phi ^2_0}{8 m R^2 \omega \Phi ^2 _{QM}} \right) \nonumber \\
    &=& \sum _{s =-\infty} ^{\infty} (-1)^{r} J_{r+2s} \left( \alpha \right) J_s \left( \beta \right) ~,
\end{eqnarray}
where we have defined the dimensionless quantities
\begin{equation}
\alpha \equiv \frac{n\hbar \Phi_0}{m R^2 \omega \Phi _{QM}} ~~~~~ \text{and} ~~~~~ \beta \equiv \frac{\hbar \Phi ^2_0}{8 m R^2 \omega \Phi ^2 _{QM}}
\label{alpha_beta}
\end{equation} 
Using this definition of $C_{r}$, the wavefunction in \eqref{wav-func-2a} becomes
\begin{equation}
    \label{wav-func-2b}
    \psi ({\bf x}, t)  = \frac{e^{i n \varphi}}{\sqrt{2 \pi R}} 
    \sum _{r=-\infty} ^{\infty} C_{r} e^{-i (r \hbar \omega + E_n ')t/ \hbar} ~.
\end{equation}
Thus, we find that when the magnetic field is turned on, the wavefunction goes from the original form in \eqref{wav-func}, to the expression shown in \eqref{wav-func-2b}. Furthermore, the original energy in \eqref{wav-func} now develops into the shifted energy $E'_n$ shown in \eqref{En'}, plus an infinite series of sidebands: $E_n' \pm r \hbar \omega$. 

We now show that \eqref{wav-func-2b} is the expected result from Floquet theory. For a sinusoidally varying Hamiltonian, as in \eqref{hamiltonian2}, Floquet theory implies that the wavefunction for the periodic Hamiltonian in \eqref{hamiltonian2} should take the general form given in equation (11) of \cite{zeldovich}
\begin{equation}
    \label{floquet}
    \psi (x,t) = \sum _k c_k \varphi (x, t) e^{-i F_k t}~.
\end{equation}
Comparing \eqref{wav-func-2b} with \eqref{floquet} we find the following correspondences: $c_k \to C_r$, $\varphi (x, t) \to \frac{e^{in\varphi}}{\sqrt{2 \pi R}}$ and $F_k \to E_n ' + r \hbar \omega$. The wavefunction in \eqref{floquet} is a superposition of quasi-states, or eigenstates of the Floquet operator, characterized by $e^{-i F_k t}$. Reference \cite{zeldovich} refers to $F_k$ as the quasi-energies. Thus, from \eqref{wav-func-2b} we find that $E_n' + r \hbar \omega$ are the quasi-energies of our system. References \cite{DK} \cite{zeldovich} make the point that transitions between these quasi-energies are observable.

Each quasi-energy comes with a weighting factor $C_r$ given by $\eqref{Ck}$. The complicated structure of $C_{r}$ makes it difficult to physically interpret the general result in $\eqref{wav-func-2b}$. However, the analysis simplifies considerably for the ground state $n=0$. Thus, we will first look at the special cases $n=0$ and $n=1$ in the following two subsections.

\subsection{The $n=0$ case}
With $n=0$, the first Bessel function in \eqref{Ck} becomes $J_{r+2s} (0)$ which is zero unless $r+2s =0$ or $s = -r/2$, which in turn means that $r$ must be even. Therefore, the weighting factor in \eqref{Ck} is $C_r = 0$ for odd values of the index $r$. For even values of $r$, the weighting factor becomes
\begin{equation}
    \label{Ck2}
    C_{r} =  (-1)^r  J_{-r/2} ( \beta ) ~~~~\to ~~~~  (-1)^{r/2}  J_{r/2} ( \beta ) ~,
\end{equation}
 where in the last step, we have used $J_{-r/2}=(-1)^{r/2} J_{r/2}$ and $(-1)^{r}=1$, since $r$ is even. Going from \eqref{Ck} to \eqref{Ck2}, there is no longer a need for the summation over $s$ since $J_{r+2s} (0) =0$ except for $s=-r/2$. Using \eqref{Ck2} in \eqref{wav-func-2b},  the wavefunction for the ground state $n=0$ becomes
\begin{equation}
    \label{wav-func-2c}
    \psi _{n=0} ({\bf x}, t)  = \frac{1}{\sqrt{2 \pi R}} 
    \sum _{r \text{ even}} (-1)^{r/2}  J_{r/2} ( \beta ) e^{-i (r \hbar \omega + E_0 ')t/ \hbar} ~.
\end{equation}

\begin{figure}[ht]
 \centering
\includegraphics[width=182mm]{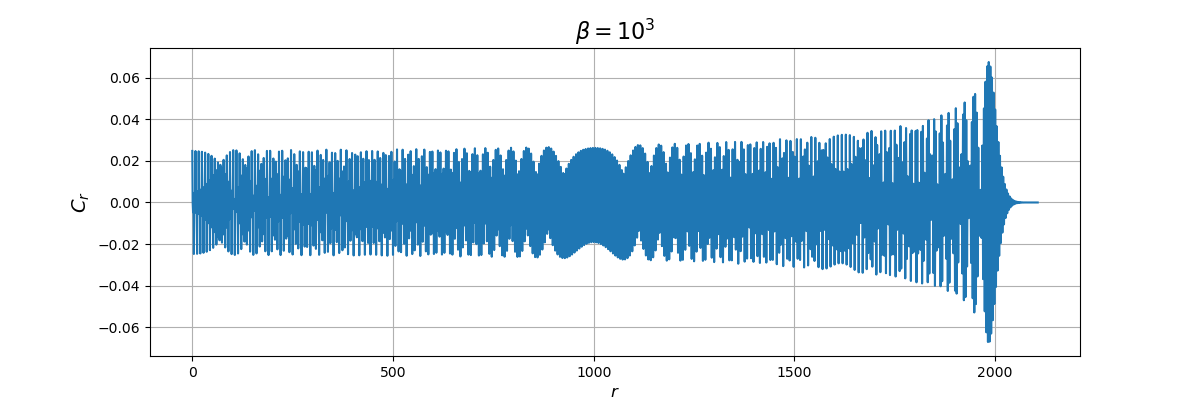}
 \caption{For $n=0$, the weighting function $C_r (\beta)$ versus the index $r$ for $\beta = 10^3$.}
\label{fig1}
\end{figure}

\begin{figure}[ht]
 \centering
\includegraphics[width=182mm]{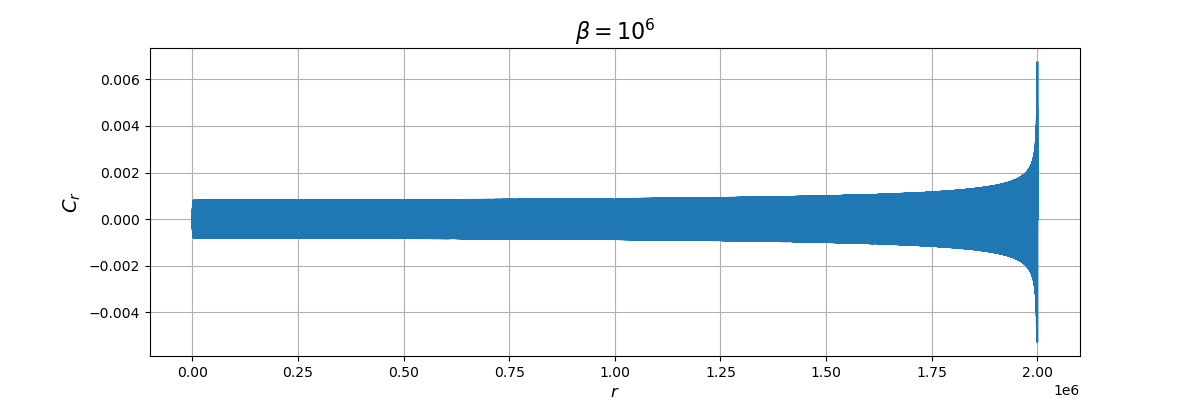}
 \caption{For $n=0$, the weighting function $C_r (\beta)$ versus the index $r$ for $\beta = 10^6$.}
\label{fig2}
\end{figure}

It appears that there are an infinite number of sidebands for the ground state energy given by $E_0 ' + r \hbar \omega$. However, the Bessel function weighting factor, $J_{r/2} ( \beta)$, cuts off the sum once $|r|$ becomes large. To see this effect, we plot the weighting factor $C_r$ as a function of the index $r$, for a fixed parameter $\beta = 10^3$ in Fig. \eqref{fig1}, and $\beta = 10^6$ in Fig. \eqref{fig2}. 

Fig. \eqref{fig1} and Fig. \eqref{fig2} both show that although the $n=0$ wavefunction of \eqref{wav-func-2c} has an infinite number of terms and energy sidebands ({\it i.e.} $E'_0+ r \hbar \omega $ with the index $r$ ranging from $-\infty$ to $+\infty$), the weighting factor $C_r(\beta)$ strongly favors a positive and negative value of $r$,  and these values are approximately given by $r_{peak} \approx 2 \beta$, where $r_{peak}$ is defined as the positive index which corresponds to the weighting factor $C_r$ with the greatest magnitude. Thus, the parameter $\beta$ determines the maximum $r$ as $r_{peak} \approx 2 \beta$. In addition, Figs. \eqref{fig1} and \eqref{fig2} show that as $\beta$ increases, the weighting of these states becomes relatively larger compared to the values of $|r| < r_{peak}$. However, the absolute value of this weighting at $r_{peak}$ becomes smaller as $\beta$ increases. It is these most prominent energy sidebands at
\begin{equation}
    \label{sb-1}
    E_0' \pm r_{peak} \hbar \omega ~~\to~~  \frac{\hbar ^2 }{2mR^2} \left( \frac{\Phi^2 _0}{2 \Phi_{QM} ^2} \right) \pm 2 \beta \hbar \omega ~,
\end{equation}
which are the signatures for this effect. 

\subsection{The excited states case ($n \ne 0$)}

In this subsection, we consider the case where $n \ne 0$. In particular, we take $n=1$ as it demonstrates the complications that generally occur for $n \ne 0$. From equations \eqref{wav-func-2a} and \eqref{Ck}, the wavefunction and coefficients become more complicated relative to the $n=0$ case. However, we can still find values of the parameters for which the wavefunction coefficients, $C_r$, have a behavior similar to that shown in Figs. \eqref{fig1} and \eqref{fig2} for the $n=0$ case, namely, $C_r$ will have a large peak for a value $r =\pm r_{peak}$, indicating that this term in the wavefunction sum in \eqref{wav-func-2b} will be the dominant term. 

\begin{figure}[ht]
 \centering
\includegraphics[width=180mm]{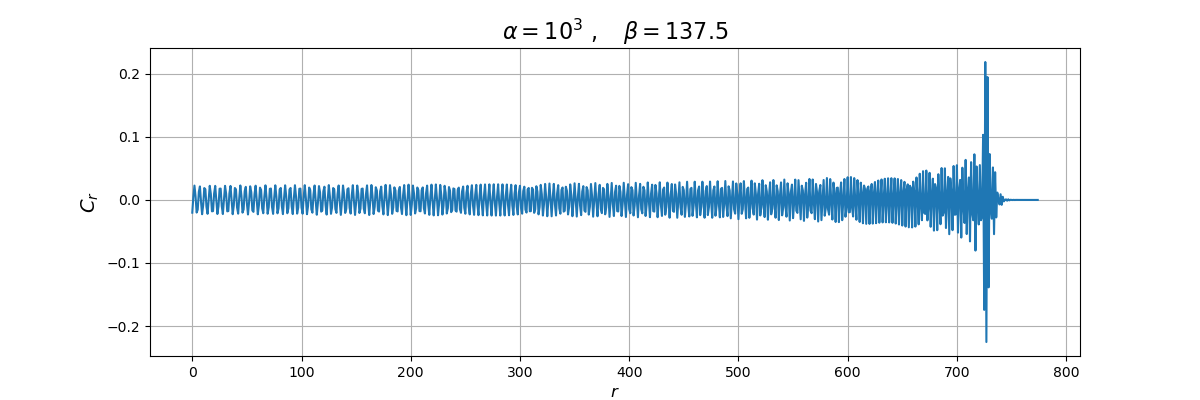}
 \caption{Plot of the weighting coefficient $C_r$ from \eqref{Ck} as a function of the index $r$ for $n=1$. For this plot, $\alpha =10^3$ and the flux ratio was chosen as $\frac{\Phi_0}{\Phi_{QM}} =1.1$ which gave $\beta = \frac{\Phi_0}{\Phi_{QM}} \frac{\alpha}{8n}= 137.5$.}
\label{fig3}
\end{figure}

\begin{figure}[ht]
 \centering
\includegraphics[width=180mm]{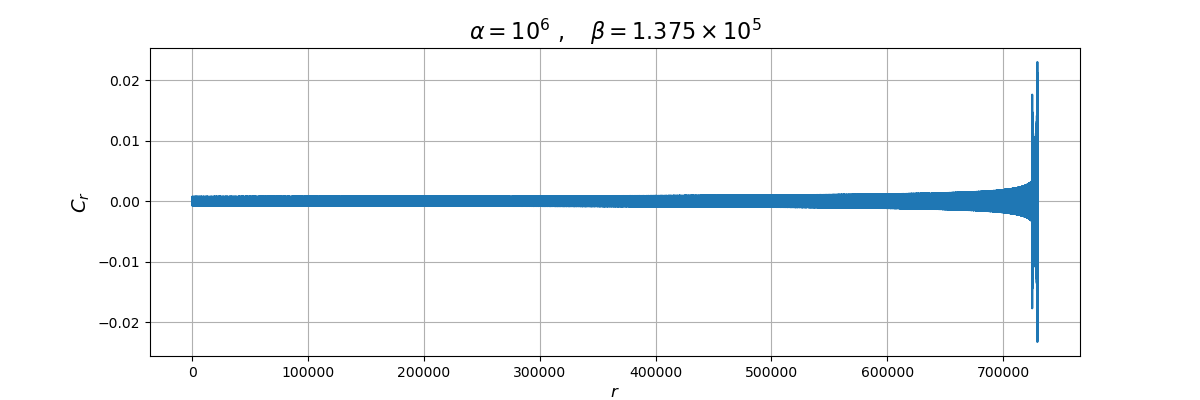}
 \caption{Plot of the weighting coefficient $C_r$ from \eqref{Ck} as a function of the index $r$ for $n=1$. For this plot $\alpha = 10 ^6$, and we choose the flux ratio $\frac{\Phi_0}{\Phi_{QM}} =1.1$ which gave $\beta = \frac{\Phi_0}{\Phi_{QM}} \frac{\alpha}{8n} = 1.375 \times 10^5$.}
\label{fig4}
\end{figure}

In calculating $C_r$ from \eqref{Ck},  we truncated the series at some finite value of $|s|_{max}$ by approximating $C_r$ via the expression 
\begin{equation}
C_r \approx \sum _{s =- s_{max}} ^{s_{max}} (-1)^{r} J_{r+2s} \left( \alpha \right) J_s \left( \beta \right)~ .
\end{equation}
This was done by numerically evaluating the series in \eqref{Ck} for increasing values of the summation index $|s|$, and stopping once the values of $C_r$ changed by less than $10^{-12}$. In Figs. \eqref{fig3} and \eqref{fig4}, we plotted the coefficients $C_r$ for certain values of $\alpha$ and $\beta$.  For $n \ne 0$, both $\alpha$ and $\beta$ are non-zero, and we choose to set $\alpha$ and $\frac{\Phi_0}{\Phi_{QM}}$ to the values listed in Figs. \eqref{fig3} and \eqref{fig4} to give values of $\beta = \frac{\Phi_0}{\Phi_{QM}} \frac{\alpha}{8n}$.

For the choice of parameters $\alpha = 10^3$, a flux ratio of $\frac{\Phi_0}{\Phi_{QM}} =1.1$, giving $\beta = \frac{\Phi_0}{\Phi_{QM}} \frac{\alpha}{8n} = 137.5$, the plot of $C_r$ versus $r$ is shown in Fig. \eqref{fig3}. Comparing this $n=1$ case with the $n=0$ case for $\beta = 10^3$ in Fig. \eqref{fig1} shows that the two cases have qualitatively similar shapes. The weighting factors in Fig. \eqref{fig1} and Fig. \eqref{fig3} increase greatly when $r = r_{peak}$. These large weighting factors imply that the wavefunctions are dominated by the value at $r_{peak}$. However, there is a quantitative difference between the weighting factors for $n=0$ and $n=1$. The value of the peak near $r = r_{peak}$ in Fig. \eqref{fig3} is much larger (at a value of $C_r \approx 0.2$) than the peak in Fig. \eqref{fig1} (at a value of $C_r \approx 0.06$). This indicates that the effect may be easier to see for the $n \ne 0$ case.
For the parameter $\beta \approx 10^6$, the comparison between the two cases is shown in Fig. \eqref{fig4} for the case $n=1$, and Fig. \eqref{fig2} for the case $n=0$. Similar comments apply to the comparison between Figs. \eqref{fig2} and \eqref{fig4}: both have the same general shape, indicating that the wavefunctions are dominated by $r_{peak}$; Fig. \eqref{fig4} has a much larger value, both in absolute terms and in relative terms, for $C_{r_{peak}}$ as compared to $C_{r_{peak}}$ in Fig. \eqref{fig2}.

As for the $n=0$ case, the signatures for this effect are the sidebands
\begin{equation}
    \label{sb-2}
    E_1' \pm r_{peak} \hbar \omega ~~\to ~~ \frac{\hbar ^2 }{2mR^2} \left( 1+ \frac{\Phi^2 _0}{2 \Phi_{QM} ^2} \right) \pm r_{peak} \hbar \omega
\end{equation}
which are spectroscopically detectable. 

 \begin{figure}[ht]
 \centering
\includegraphics[width=150mm]{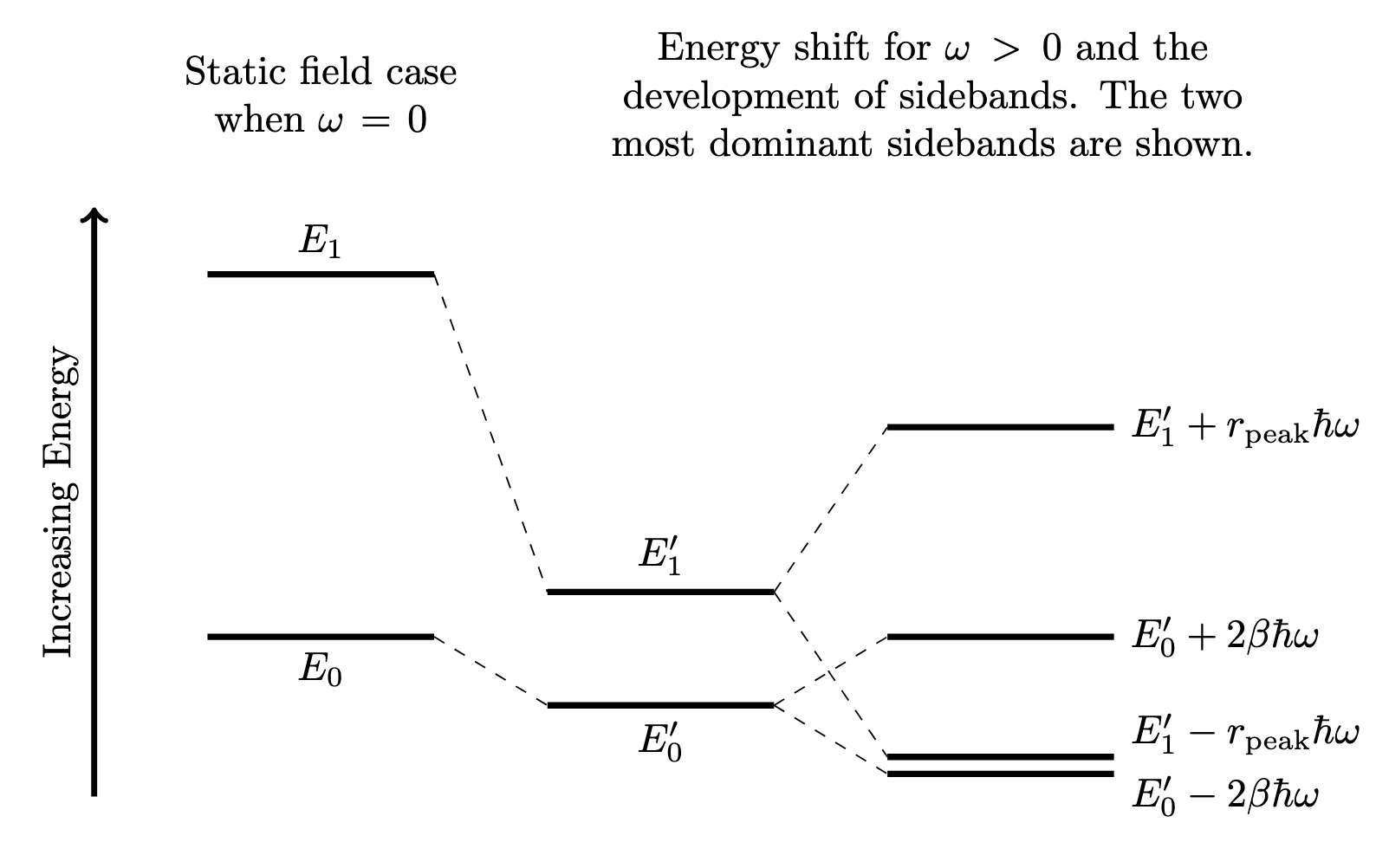}
 \caption{The energy level diagram for the $n=0$ and $n=1$ states for the static magnetic field (far left in diagram) and then the shifting of the energies into quasi-energy sidebands (far right in the diagram). Only the most prominent sidebands are shown.}
\label{levels}
\end{figure}

The energy level diagram in Fig. \eqref{levels} illustrates the splitting of the $n = 0$ and $n = 1$ states into quasi-energy sidebands. For each case, the two most prominent sidebands are shown, corresponding to $\pm r_{peak}$. In actuality, the full spectrum of quasi-energies for $|r| < r_{peak}$ is present in principle.

As shown in the diagram, the lower sideband of the $n = 1$ state, $E_1' - r_{peak} \hbar \omega$, drops below the upper sideband of the $n = 0$ state, $E_0' + 2 \beta \hbar \omega$. However, this crossing does not violate the no-crossing theorem \cite{no-cross}, since the two states involved are orthogonal, belonging to different states of angular momentum. Thus, the crossing is physically allowed.

It is also noteworthy that the lower sidebands for $n = 0$ and $n = 1$ nearly coincide. This near-alignment is an artifact of the specific values chosen for $\alpha$, $\beta$, and $\Phi_0$. The size of the gap between these levels can therefore be adjusted by adjusting the values of these parameters. However, the gap between $E_1'- r_{peak} \hbar \omega$ and $E_0'-2 \beta \hbar \omega$ is smaller than the gap between other levels. 

In principle, transitions between all the sidebands on the right in Fig. \eqref{levels} can occur, but some transitions are suppressed. For example, a transition between the upper and lower sidebands of $E_1 '$ would have a change in the angular momentum of $\Delta l =0$, which is suppressed relative to transitions with $\Delta l = \pm 1$. Thus, it is more likely one will see transitions from the two $E'_1$ sidebands to the two $E'_0$ sidebands.

\subsection{Persistent Currents} In the above, we have focused on the shifting of the energy levels of the quantum ring to probe the effect of the sinusoidal varying magnetic flux. However, the system of a quantum ring threaded by a constant magnetic flux also results in a persistent current in the ring. In this subsection, we will see to what extent this persistent current carries over to the case of a sinusoidally varying magnetic flux. 

The current density in the ring can be determined using the wavefunction from \eqref{wav-func-2} and the vector potential in \eqref{vec-pot2} to calculate the probability current, ${\bf J}_{(prob)}$. The current density is then the probability current multiplied by the charge, $e$
\begin{equation}
    \label{current-den}
    {\bf J} = e {\bf J}_{(prob)}=\frac{e}{2m}  \left( \Psi^* {\bf D} \Psi - \Psi ({\bf D } \Psi )^*\right)~,
\end{equation}
where ${\bf D} \Psi = -i \hbar \nabla \Psi - e {\bf A} \Psi$ and  $({\bf D} \Psi)^* = -i \hbar \nabla \Psi^* + e {\bf A} \Psi^*$. Due to the form of $\Psi$ from \eqref{wav-func-2} and since ${\bf A} \propto {\hat \varphi}$ we have ${\bf J}_{(prob)} \propto {\hat \varphi}$ {\it i.e.} the current density is only in the $\varphi$-direction. In detail, using \eqref{wav-func-2} and  \eqref{vec-pot2} in \eqref{current-den} we obtain
\begin{eqnarray}
    \label{current-den1}
     J_{(prob)}^{\varphi} &=& -\frac{i e \hbar}{2m} \left(\Psi^* \frac{1}{R}\frac{\partial \Psi}{ \partial \varphi} -\Psi \frac{1}{R}\frac{\partial \Psi^*}{ \partial \varphi} \right) - \frac{e^2 }{m}\Psi^* \Psi A_\varphi\nonumber \\
     &=&\frac{en\hbar}{2 \pi m R^2} - \frac{e^2 \Phi_0}{4 \pi^2 m R^2} \cos \omega t~.
\end{eqnarray}
The first term in \eqref{current-den1} is the constant, persistent current density expected from the case of a constant magnetic flux through the ring -- see section 6 of \cite{review-qr}. The second, sinusoidal term comes from the time variation of the flux. This time-varying current density would signal the presence of the oscillating magnetic flux. From \eqref{current-den1} one sees that the current density is proportional to $\frac{1}{R^2}$, thus as the size of the ring shrinks the current density increases.  

\subsection{Continuous ring versus discrete ring}

The results presented in this section are for a continuous ring. One can ask how the above results, especially the energy sidebands, would change for an $L$-site ring, with $N$ electrons, threaded by a magnetic flux $\Phi_0$. For this situation, the Hamiltonian for the continuous ring, given in equation \eqref{hamiltonian2}, is replaced by the Hamiltonian of the Hubbard model \cite{review-qr}
\begin{eqnarray}
    \label{hubbard}
    H_{Hubbard}= -t \sum_{i=1} ^N \sum _{\sigma} \left( e^{-i2 \pi \phi/L}c^{\dagger}_{i+1, \sigma} c_{i,\sigma} + e^{i2 \pi \phi/L}c^{\dagger}_{i, \sigma} c_{i+1,\sigma}\right) + U \sum_{i=1} ^N {\hat n}_{i, \uparrow}{\hat n}_{i, \downarrow} ~,
\end{eqnarray}
where $\phi = \frac{\Phi_0}{\Phi_{QM}}$ is a ratio of magnetic fluxes,  $c^{\dagger}_{i, \sigma}$ ($c_{i,\sigma}$) is the creation (annihilation) operator for an electron of spin $\sigma$ at site $i$, ${\hat n}_{i, \sigma} = c^{\dagger}_{i, \sigma}c_{i, \sigma}$ is the number operator for a spin-$\sigma$ electron at site $i$, $t$ is a hopping parameter, and $U$ is an interaction energy. The first, double sum term ({\it i.e.} the ``kinetic term") in \eqref{hubbard}  describes the hopping of electrons between neighboring sites, and the second, single sum term (``potential term") describes the repulsion between electrons at the same site.   

As discussed in \cite{review-qr} (see in particular section 9), the general analysis of the system using $H_{Hubbard}$ for a large $N$ and $L$ is complicated, and must be done numerically, via computer simulations. However, in the limit of a single electron, $N=1$, and a large number of sites, $L \to \infty$, the results of using $H_{Hubbard}$ approach those of using the Hamiltonian for the continuous ring given in equation \eqref{hamiltonian2}. 

\section{Possible experimental realization}

In this section, we discuss, broadly, what parameters would be reasonably accessible in an experiment to test the general predictions for this system of a quantum ring threaded by a sinusoidally varying magnetic flux. The parameters from the analysis of section II are: (i) $R$, the radius of the quantum ring; (ii) $\omega$, the frequency at which the magnetic flux is varied; (iii) $\Phi_0$, the magnitude of the magnetic flux; (iv) $m$, the mass of the quantum particle confined to the ring. 

We used the review article on quantum rings \cite{review-qr} to inform our choices for the parameters. First, most quantum rings have electrons, as the quantum system is confined to the ring. This fixes our mass at $m= 9.11 \times 10^{-31}$ kg.  Next, from \cite{review-qr} the radius of quantum rings has been decreasing over time from micron size to about a hundred times nanometer size. Thus, we take the ring radius in the range $10^{-7} ~ {\rm m} \le R \le 10^{-3} ~ {\rm m}$. We have taken the upper limit on $R$ to be $10^{-3}$ m since, as we shall see shortly, larger $R$ can more easily accommodate the theoretically optimal ranges of $\alpha$ and $\beta$ from Figs. \eqref{fig1}-\eqref{fig4}. Additionally, a larger quantum ring is easier to construct and easier to thread a solenoid through.

As mentioned in section II, the approximations used on the vector potential in \eqref{vec-pot} required $kR \ll 1 \to \omega R \ll c$. Given the smallness of $R$, this allows flexibility in the choice of $\omega$. However,  both experimentally and theoretically, it is easier to consider frequencies in the range $10 {\rm ~Hz} \le \omega \le 1000 {\rm ~Hz}$. A larger $\omega$ would require taking into account radiation from the solenoid. In any case with these ranges for $R$ and $\omega$ the condition, $\omega R \ll c$, is met.

Both parameters $\alpha$ and $\beta$ in \eqref{alpha_beta} have the same factor of $\frac{\hbar}{m R^2 \omega}$. The parameter $\alpha$ has two additional multiplicative factors: $n$ and $\frac{\Phi_0}{\Phi_{QM}}$. Similarly, the parameter $\beta$ has multiplicative factors: $\frac{1}{8}$ and $\left(\frac{\Phi_0}{\Phi_{QM}} \right)^2$.  
Given a frequency range of $10 {\rm ~Hz} \le \omega \le 1000 {\rm ~ Hz}$, and a range of ring radius, $10^{-7} {\rm ~ m}\le R \le 10^{-3} {\rm ~ m}$,  we find $10^{-13} {\rm ~\frac{m^2}{s}} \le \omega R^2 \le 10^{-3} {\rm ~\frac{m^2}{s}}$. Next, assuming that the flux $\Phi _0$ takes approximately the minimum, non-zero value of $\Phi _{QM}$, and using $n=1$, leads to the following bounds on $\beta \approx \frac{\hbar}{8m R^2 \omega}$ and $\alpha \approx \frac{\hbar}{m R^2 \omega}$ given as 
\begin{equation}
    \label{ab}
    1.45 \times 10^{-2} \le \beta \le 1.45 \times 10^8 ~~~~~{\rm and}~~~~~1.16 \times 10^{-1} \le \alpha \le 1.16 \times 10^9 ~.
\end{equation}
From Figs. \eqref{fig1} and \eqref{fig2}, for the $n=0$ case, we see that the theoretically optimal values of $\beta$ lie in the range $10^3 \le \beta \le 10^6$  which is accommodated by the experimentally allowed range from \eqref{ab}. From Figs. \eqref{fig3} and \eqref{fig4}, for the $n=1$ case, we see that the theoretically optimal values of $\alpha$ and $\beta$, lie in the ranges $10^3 \le \alpha \le 10^6$ and $10^2 \le \beta \le 10^5$, which is accommodated by the experimentally allowed range from \eqref{ab}. 

Further comparing Figs. \eqref{fig1} \eqref{fig2} with Figs. \eqref{fig3} \eqref{fig4} leads to the following observations:
\begin{itemize}
\item For the case $n=1$, the energy sidebands, $E_1' + r_{peak} \hbar \omega$, will be more prominent than the $n=0$ sidebands, $E_0' + r_{peak} \hbar \omega$. First, the absolute value of $C_{r_{peak}}$ is larger for Figs. \eqref{fig3} \eqref{fig4} versus Figs. \eqref{fig1} \eqref{fig2}. The former have values of $C_{r_{peak}} \approx 0.21$ for Fig. \eqref{fig3} and $C_{r_{peak}} \approx 0.023$ for Fig. \eqref{fig4}, while the latter have $C_{r_{peak}} \approx 0.063$ for Fig. \eqref{fig1} and $C_{r_{peak}} \approx 0.0063$ for Fig. \eqref{fig2}. 
\item The relative size of $C_{r_{peak}}$ compared to the values of $C_r$ for $r < r_{peak}$ is greater for Figs. \eqref{fig3} \eqref{fig4} versus Figs. \eqref{fig1} \eqref{fig2}. From Fig. \eqref{fig2} the ratio of the value of $C_{r_{peak}}$ to the values of $C_r \approx 0.001$ for $r < r_{peak}$ is $\frac{C_{r_{peak}}}{C_{r}} = 6.3$; from Fig. \eqref{fig4} the ratio of the value of $C_{r_{peak}}$ to the value of $C_r \approx 0.001$ for $r < r_{peak}$ is $\frac{C_{r_{peak}}}{C_{r}} = 23$.
\end{itemize}
The overall conclusion of the above estimates is that the energy sidebands, which are the signatures of this effect, will be easier to see for larger values of $\alpha$ and $\beta$, and the $n = 1$ case will be easier to observe versus the $n=0$ case.     

There are two issues that would potentially have an effect on the experimental results: (i) disorder via impurities and (ii) temperature. The effect of disorder due to impurities was studied via numerical methods in several works \cite{impurity1,impurity2,impurity3,impurity4}. The general effect of impurities is to decrease the persistent current and to lift any degeneracy in the energy eigenvalues. The effect of temperature on quantum rings was studied using a quantum Monte Carlo code in \cite{borrmann}, where it was found that quantum rings with several electrons exhibit a transition between spin-ordered and disordered Wigner crystals, which depends on temperature, ring diameter, and particle number. It would be interesting to see how these two issues —impurities and temperature — affect the sideband structure discussed here. Such work would require numerical simulations, which is beyond the scope of the analytical work presented here.
 
\section{Summary and Conclusions}

We investigated the system of a quantum ring threaded by a sinusoidally varying magnetic field. This system is an extension of a quantum ring threaded by a static magnetic flux, which is an example of the Aharonov-Bohm effect. For the quantum ring threaded by a static magnetic flux, the signature is the shifting of energy levels given in \eqref{energy}. If the ratio of fluxes, $\frac{\Phi_0}{\Phi_{QM}}$, is an integer, then this shift is spectroscopically unobservable since the different energy levels shift into each other and the differences in energy do not change. If $\frac{\Phi_0}{\Phi_{QM}}$ is not an integer, then the energy-level shift is observable \cite{tong}. 

For a sinusoidal magnetic flux the vector potential and fields can be written down exactly; see equations \eqref{vec-pot} -- \eqref{b-fields}. Since the electric and magnetic fields from \eqref{e-fields} and \eqref{b-fields} are non-zero, this time-varying system is not an example of the Aharonov-Bohm effect. 
We have shown that this system, of a quantum ring threaded by a sinusoidally varying magnetic flux, is closely related to the ac Stark effect \cite{autler,DK}. The difference between the quantum ring plus time-varying magnetic flux versus the standard ac Stark effect is that the linear and quadratic terms for the quantum ring are directly related to each other, whereas for the standard ac Stark effect the two terms are not directly connected. For the standard ac Stark effect, the linear term comes from the interaction of the dipole moment of the material with the electric field, while the quadratic term comes from the polarizability of the material \cite{DK}. For the quantum ring plus sinusoidal magnetic flux, the linear and quadratic terms come from the same source -- the squaring of the minimal coupling ({\it i.e.} $(p_\varphi + e A_\varphi)^2$).

After taking the low-frequency limit of the vector potential in \eqref{vec-pot2}, we solved the time-dependent Schr{\"o}dinger equation exactly using the Jacobi-Anger expansion (details are in Appendix A). This led to the wavefunction in \eqref{wav-func-2b} which was an infinite sum of quasi-energy terms $e^{-i(r \hbar \omega + E_n')t/\hbar}$, with each of these terms having a weighting factor $C_r$ given in \eqref{Ck}. The general shape of $C_r$ is shown in Figs. \eqref{fig1} -- \eqref{fig4}. From these figures one can see that after some value of the summation index, $r=r_{peak}$, the weighting factor goes to zero, effectively cutting off the sum at $r=r_{peak}$. Furthermore, $C_r$ has its maximum value at $r = r_{peak}$. From \eqref{wav-func-2b} this implies that the term $\frac{e^{i n \varphi}}{\sqrt{2 \pi R}} C_{r_{peak}} e^{-i(r_{peak} \hbar \omega + E_n ')t/\hbar}$ dominates the wavefunction, and that the energy $E_n '$ from \eqref{En'} develops dominant quasi-energy sidebands at $\pm r_{peak} \hbar \omega$. These quasi-energy sidebands 
should be spectroscopically observable provided that $\frac{\Phi_0}{\Phi_{QM}}$ is not an integer.   

In section 3 we looked at the allowed values for ring radius $R$, frequency $\omega$, and magnitude of the magnetic flux, $\Phi_0$, which determined how observable the energy sidebands would be. The general conclusion was that values of $R, \omega, \Phi_0$ which gave larger values of $\alpha$ and $\beta$ were better, and that $n =1$ was more observable compared to $n=0$.  \\

{\bf Acknowledgments:} DS acknowledges the Frank Sutton Research Fund for support during the completion of this work.

\appendix 

\section{Jacobi-Anger analysis of sinusoidal term in \eqref{wav-func-2}}

The two sinusoidal terms in \eqref{wav-func-2} can be handled {\it individually} using the Jacobi-Anger expansion
\begin{equation}
    \label{ja}
    e^{-iz \sin (\theta )} = \sum _{r=-\infty} ^{\infty} J_r (z) e^{-i r \theta} ~,
\end{equation}
where $J_r (z)$ are $r^{th}$ order Bessel functions. The Jacobi-Anger expansion of the $\sin( \omega t)$ term in $\eqref{wav-func-2}$ gives 
\begin{equation}
    \label{ja-1}
    \exp\left(-i \frac{\hbar n \Phi_0 }{mR^2 \omega \Phi_{QM}}\sin (\omega t)\right) = \sum _{r=-\infty} ^{\infty} J_r \left( \frac{n\hbar \Phi_0}{m R^2 \omega \Phi _{QM}} \right) e^{-ir \omega t} ~,
\end{equation}
For the $\sin( 2 \omega t)$ term in \eqref{wav-func-2}, the Jacobi-Anger expansion gives
\begin{equation}
    \label{ja-2}
    \exp\left(-i \frac{\hbar \Phi_0 ^2}{8 m R^2 \omega \Phi ^2_{QM}}\sin (2 \omega t)\right) = \sum _{s=-\infty} ^{\infty} J_s \left( \frac{\hbar \Phi ^2_0}{8 m R^2 \omega \Phi ^2 _{QM}} \right) e^{-i 2 s \omega t}
\end{equation}
Now we let $r \to -r$ in \eqref{ja-1} and combine this with \eqref{ja-2} to yield
\begin{equation}
    \label{ja-3}
\sum _{r=-\infty} ^{\infty} \sum _{s=-\infty} ^{\infty}  J_{-r} \left( \frac{n\hbar \Phi_0}{m R^2 \omega \Phi _{QM}} \right) J_s \left( \frac{\hbar \Phi ^2_0}{8 m R^2 \omega \Phi ^2 _{QM}} \right) e^{i (r-2 s) \omega t} ~.
\end{equation}
Now we shift the $r$ index as $r \to  r+ 2s$ and then re-write \eqref{ja-3} as
\begin{equation}
    \label{ja-4}
\sum _{r = -\infty} ^{\infty} \sum _{s = -\infty} ^{\infty}  J_{-r-2s} \left( \frac{n\hbar \Phi_0}{m R^2 \omega \Phi _{QM}} \right) J_s \left( \frac{\hbar \Phi ^2_0}{8 m R^2 \omega \Phi ^2 _{QM}} \right) e^{i r \omega t} ~,
\end{equation}
Now using the identity $J_{-n} = (-1)^n J_n$, we can re-write the first Bessel function in \eqref{ja-4} as $J_{-r-2s} = (-1)^{r+2s} J_{r+2s} = (-1)^{r} J_{r+2s}$ since $(-1)^{2s} =1$. Finally, using the fact that the Hamiltonian and the time-dependent Schr{\"o}dinger equation ($i \hbar \partial_t \Psi = H \Psi$) are invariant under $t \to -t$, we re-write \eqref{ja-3} as
\begin{equation}
    \label{ja-5}
\sum _{r =-\infty} ^{\infty}  \sum _{s =-\infty} ^{\infty}  (-1)^{r} J_{r+2s} \left( \frac{n\hbar \Phi_0}{m R^2 \omega \Phi _{QM}} \right) J_s \left( \frac{\hbar \Phi ^2_0}{8 m R^2 \omega \Phi ^2 _{QM}} \right) e^{-i r \omega t} \equiv \sum _{r=-\infty} ^{\infty}  C_{r} e^{-i r \omega t} ~.
\end{equation}
In the last step in \eqref{ja-5} we have defined $C_{r}$ as
\begin{equation}
    \label{ja-6}
C_{r} \equiv \sum _{s =-\infty} ^{\infty}  (-1)^{r} J_{r+2s} \left( \frac{n\hbar \Phi_0}{m R^2 \omega \Phi _{QM}} \right) J_s \left( \frac{\hbar \Phi ^2_0}{8 m R^2 \omega \Phi ^2 _{QM}} \right)  ~.
\end{equation}
The result for $C_{r}$ is equivalent to equation (17) of reference \cite{DK} which we repeat here for comparison
\begin{equation}
    \label{ja-7}
C_{r} \equiv \sum _{s =-\infty} ^{\infty}  (-1)^{r} J_{r+2s} \left( \frac{d F}{\hbar \omega} \right) J_s \left( \frac{ \alpha_0 F ^2}{8 \hbar \omega} \right)  ~.
\end{equation}
where $F$ is the electric field strength, $d$ is the constant dipole moment, and $\alpha_0$ is the polarizability. Reference \cite{DK} sets $\hbar =1$, however, in \eqref{ja-7} we have restored $\hbar$, as well as changing the index in \cite{DK} from $k$ to $r$.  Comparing the mathematical form of $\eqref{ja-6}$ and $\eqref{ja-7}$, it is evident that the electric field magnitude is replaced by the ratio of magnetic fluxes: $F \leftrightarrow \frac{\Phi_0}{\Phi_{QM}}$. The polarizability is replaced by $\alpha_0 \leftrightarrow \frac{\hbar^2}{m R^2}$, and the dipole moment is replaced by $d \leftrightarrow \frac{n \hbar ^2 }{m R^2}$. While the result from \cite{DK} given in \eqref{ja-7} is mathematically similar to our result in \eqref{ja-6}, the physical basis for the results are different. In \eqref{ja-7} the linear term, $\frac{d F}{\hbar \omega}$, comes from the interaction of the electric field strength $F$ with the dipole moment $d$, while the quadratic term, $\frac{ \alpha_0 F ^2}{8 \hbar \omega}$,  is the interaction of the electric field strength $F$ with the material polarizability $\alpha_0$. Therefore, the linear and quadratic terms in \eqref{ja-7} can be independent based on the particular material considered. In contrast, the linear and quadratic terms in $\eqref{ja-6}$ arise from the minimal coupling of the charged particle to the vector potential in \eqref{hamiltonian}  which involves $(p_\phi + e A_\phi)^2$ in the Hamiltonian. Therefore, unlike the case in \eqref{ja-7}, the linear and quadratic terms in \eqref{ja-6} cannot be adjusted independently. They are necessarily linked.

\end{document}